\documentclass[useAMS,usenatbib,usegraphicx]{mn2e}
\usepackage{rotating}
\usepackage{txfonts}

\def\lae{\mathrel{<\kern-1.0em\lower0.9ex\hbox{$\sim$}}}
\def\gae{\mathrel{>\kern-1.0em\lower0.9ex\hbox{$\sim$}}}

\title[The assembly of normal and compact ETGs]
{Constraining the star formation and the assembly histories of normal and 
compact early-type galaxies at $1<z<2$.}
\author[P. Saracco et al.]{P. Saracco$^{1}$\thanks{E-mail: 
paolo.saracco@brera.inaf.it}, M. Longhetti$^{1}$,  A. Gargiulo$^{1}$\\ 
$^{1}$INAF - Osservatorio Astronomico di Brera, Via Brera 28, 20121 Milano,
Italy
}
\begin{document}

\date{Accepted 2010 November 25. Received 2010 November 11; in original form
2010 August 11}
\pagerange{\pageref{firstpage}--\pageref{lastpage}} \pubyear{2010}
\maketitle
\label{firstpage}
\begin{abstract}
We present a study based on a sample of 62 early-type galaxies (ETGs)
at $0.9<z_{spec}<2$ aimed at constraining their past
star formation and mass assembly histories.
The sample is composed of normal ETGs having effective radii comparable 
to the mean radius of local ones and of compact ETGs having effective radii 
from two to six times smaller.
We do not find evidence of a dependence of the compactness
of ETGs on their stellar mass.
The best fitting to their spectral energy distribution at known redshift
has allowed us to constrain the epoch at which the stellar mass formed.
We find that the stellar mass of normal ETGs formed at $z_{form}\lae3$ 
while the stellar content of compact ETGs formed  over a wider
range of redshift ($2<z_{form}<10$) with a large fraction of them 
characterized by $z_{form}>5$.
Earlier stars, those formed at $z_{form}>5$,
are assembled in compact and more massive ($\mathcal{M}_*>10^{11}$
M$_\odot$) ETGs while stars later formed ($z_{form}\lae3$) or resulting
from subsequent episodes of star formation are assembled
both in compact and normal ETGs. 
Thus, the older the stellar population the higher the 
mass of the hosting galaxy but not vice versa.
This suggests that the epoch of formation 
may play a role in the formation of massive ETGs rather than the mass itself.
We show that the possible general scheme in which normal ETGs at 
$\langle z\rangle\simeq1.5$ are descendants of compact spheroids 
assembled at higher redshift is not compatible with the current models.
Indeed, we find that the number of dry mergers expected in a hierarchical
model is almost two orders of magnitude lower than the one needed to
enlarge a compact ETGs up to a normal-size ETG. 
Moreover, we do not find evidence supporting a dependence of the compactness
of galaxies on their redshift of assembly, a dependence expected  
in the hypothesis that the compactness of a galaxy is due to  the higher 
density of the Universe at earlier epochs.
Finally, we propose a simple scheme of formation and assembly of the stellar 
mass of ETGs based on dissipative gas-rich merger which can qualitatively 
account for the co-existence of normal and compact ETGs observed at 
$\langle z\rangle\simeq1.5$ in spite of the same stellar mass, the lack of 
normal ETGs with high $z_{form}$ and the absence of correlation between 
compactness, stellar mass and formation redshift.
\end{abstract}


\section{Introduction}
To understand how galaxies have formed and evolved it is fundamental to know
which and how many of them were present in the Universe at different epochs. 
In the last years, many efforts have been devoted to trace the evolution of 
early-type galaxies back in time since they contain most of the present-day 
stars and  baryons (e.g. Fukugita et al. 1998).
After the first spectroscopic detection of early-type galaxies at
$z\sim1.5$ (Dunlop et al. 1996; Saracco et al. 2003; McCarthy et al. 2004;
Cimatti et al. 2004; Glazebrook et al. 2004; Saracco et al. 2005), 
the attention was attracted by the small effective radius 
of many of them when compared to the mean radius of present-day early-types 
of comparable mass (Daddi et al. 2005; Trujillo et al. 2006; 
Longhetti et al. 2007).
Since then, many works focused on the "small early-type galaxies problem" 
corroborating the idea that in the past early-type galaxies  were 
smaller at fixed mass, hence denser
(McGrath et al. 2008; Cimatti et al. 2008; Buitrago et al. 2008; 
van Dokkum et al. 2008; Damjanov et al. 2009; Muzzin et al. 2009; 
Cassata et al. 2010; Carrasco et al. 2010).
As a consequence of this,  ETGs must have 
increased their radius across the time  to 
reconcile with the present-day early-types. 
However, evidence of a significant number of normal ETGs 
at $z\sim1.5$ following the local scaling relations has been
accumulated in the last couple of years suggesting 
that at least not all the high-z early-types were more compact 
(Saracco et al. 2009; Mancini et al. 2010) and that, consequently, not all 
the high-z ETGs undergo size evolution.
The first measurements of velocity dispersion of a few of "normal-size"
high-z ETGs confirmed that they are similar to typical local ones also 
from the dynamic point of view (Cenarro et al. 2009; Cappellari et al. 2009; 
Onodera et al. 2010).  
Concurrently, evidence of the presence  of a significant fraction of compact 
ETGs in the local Universe similar to the high-z ones came out 
(e.g. Trujillo et al. 2009; Valentinuzzi et al. 2010a; 2010b)
casting the first doubts about the size evolution scenario.
The question naturally arising from these new pieces of evidence is 
whether compact ETGs were so much more numerous at earlier epochs to require 
their effective radius evolution.
Recently, evidence that the number density of compact ETGs at 
$<z>\simeq1.5$ 
was not significantly higher than the number density of compact ETGs 
seen in local cluster of galaxies has come out (Saracco, Longhetti \&
Gargiulo 2010).
This evidence conflicts with the hypothesized effective radius evolution 
of high-z ETGs while shows that among them 
there are the progenitors of the compact ETGs seen in local 
clusters of galaxies and that they  were  as we see them today
already 9-10 Gyr ago as confirmed by recent studies on high-z cluster
galaxies (e.g. Strazzullo et al. 2010).
Moreover, at $z\sim1.5$  a majority of normal ETGs 
co-exist with compact early-types from $\sim2$ to $\sim6$ times smaller 
in spite of the same mass and redshift.
Actually, this picture is not different at least qualitatively 
from what is observed in the local universe: most of the ETGs lie on a 
well defined scaling relation and a minor fraction of them 
(e.g. $\sim$20-40 per cent  in cluster of galaxies, Valentinuzzi et al. 
2010a; 2010b) are significantly denser than the others.
Thus, ETGs appear a composite population from $z=0$ up to at least 
$z\sim1.5-2$.
To corroborate this view is the recent study conducted by Gargiulo et al. 
(2010) who show that at $1<z<2$ ETGs with negative color gradient 
(redder toward the center) co-exist with ETGs characterized by positive
color gradient (bluer toward the center).
Consequently, this non homogeneity of the population of ETGs must originate 
at an earlier epoch ($z>2$) when they have been assembled.
The relevant question is which formation scenario and early physical 
conditions can account for the observed different properties of ETGs.
In this paper we try to constrain these issues by probing 
the past history of a large number of ETGs at  $0.9<z_{spec}<2$. 
In Sec. 2 we describe the data set we used in this analysis.
In Sec. 3 we describe our analysis and we present the results.
In Sec. 4 we use the results obtained to derive some constraints on the
spheroids formation at very early epochs.
Finally, in Sec. 5, we summarize the results and present the conclusions.
Throughout this paper we use a standard cosmology with
$H_0=70$ Km s$^{-1}$ Mpc$^{-1}$, $\Omega_m=0.3$ and $\Omega_\Lambda=0.7$.
All the magnitudes are in the Vega system, unless otherwise specified.

\section{The data set}
The sample of ETGs we used in our analysis is composed of 62 ETGs
in the spectroscopic redshift range $0.9<z_{spec}<2$ and  with
magnitudes in the range $17<K<20.5$. 
The whole sample is covered by HST observations at 
spatial resolution of about 0.8 kpc (FWHM$\sim~0.1$ arcsec) at the 
redshift of the galaxies. 
Out of the 62 ETGs, 28 are covered by observations with the Near Infrared 
Camera and Multi Object Spectrograph (NICMOS sample hereafter) in the F160W 
filter and 34 are covered by observations with the Advanced Camera for 
Surveys (ACS sample hereafter). 
The NICMOS sample  is a compilation of early-type 
galaxies at $1.2<z_{spec}<1.85$ which combines proprietary  data 
(Longhetti et al. 2007) with those from other surveys for a total of 32 ETGs
in its original form (Saracco et al. 2009). 
Here, we removed 4 galaxies due to the poor fitting to 
their profile obtained on the NIC3 images.
The wavelength coverage is composed of 10 photometric bands 
(from 0.4 µm to 3.6 µm) for most of the sample. 
The ACS sample  is a complete sample of ETGs 
we selected at K$\leq20.2$ (Saracco et al. 2010) on the southern field of the  Great 
Observatories Origins Deep Survey (GOODS; Giavalisco et al. 2004) 
with redshift $0.9<z_{spec}<1.92$. 
The wavelength coverage of the ACS sample is composed of  14 photometric 
bands extending from 0.3 $\mu$m to 8.0 $\mu$m. 
The complete ACS sample will be presented in a forthcoming paper
(Saracco et al. in preparation).
   
The effective radii of the galaxies were derived by fitting a 
S\'ersic profile to the observed light profile in the HST
NICMOS-F160W and ACS-F850LP images. 
Stellar masses $\mathcal{M}_*$ and ages of the stellar populations were
derived by fitting the red-shifted templates to the observed spectral energy
distribution of the galaxy.
We used the last release of the stellar population synthesis models 
of Charlot \& Bruzual with Chabrier initial mass function 
(IMF, Chabrier 2003), four exponentially declining star formation histories 
(SFHs) with e-folding time $\tau=[0.1, 0.3, 0.4, 0.6]$ Gyr and metallicity
0.4 $Z_\odot$, $Z_\odot$ and 2 $Z_\odot$.
The profile fitting and the SED fitting of galaxies are described 
in detail in the previous papers (Saracco et al. 2009; 2010).
{ In Tab. 1 we report the main properties and the best fitting parameters 
for the 62 ETGs considered in the present analysis.}

\begin{figure}
\begin{center}
\includegraphics[width=8.6cm]{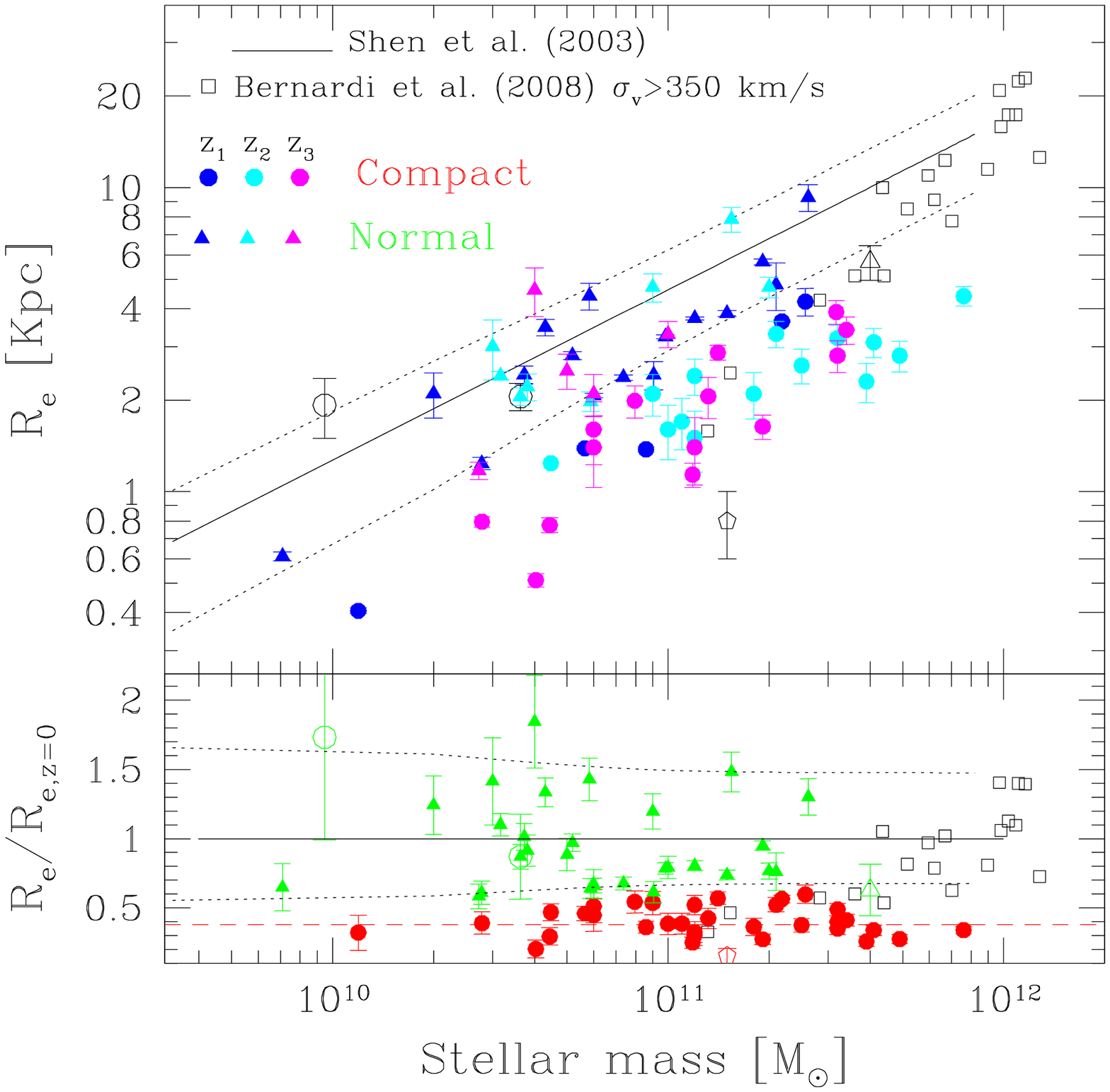} 
\vskip -0.5truecm
\caption{Upper panel - Effective radius R$_e$ versus stellar mass for
our sample of 62 ETGs.
The solid line is the local Size-Mass relation  (Shen et al. 2003). 
Normal ETGs lying within  1 sigma (dotted lines) from the local relation, 
are marked by filled triangles. 
Compact early-types diverging more than 1$\sigma$ from the local relation 
{ (including errors on size measurements)} are marked by filled circles.
{ Filled symbols are color-coded according to the redshift of the galaxy.
In particular, the three different colors blue, cyan and magenta 
mark the three redshift intervals $0.9<z_1<1.2$, $1.2<z_2<1.5$
and  $1.5<z_3<2$ respectively. }
The two open circles represent the ETGs with velocity dispersion
measured by Cappelari et al. (2009)
one of which belonging to our ACS sample,
the open pentagon is the one measured by van Dokkum et al. (2009)
and the open triangle is the one by Onodera et al. (2010).
The open squares represents the 29 early-types selected by Bernardi et al. 
(2008) at $z<0.1$ with velocity dispersion $\sigma_v>350$ km/s. 
Lower panel - The compactness, defined as the ratio between the 
effective radius R$_e$ of the galaxy and the effective radius R$_{e,z=0}$
of an equal mass galaxy at $z=0$ as derived by the local S-M relation, 
is plotted as a function of the stellar mass. Symbols are as in the upper
panel. In this case normal ETGs  are marked by green filled triangles
while compact ETGs by red filled circles for clarity. The (red) dashed line
represents the mean compactness value of compact ETGs
R$_e$/R$_{e,z=0}\simeq0.38$. 
}
\end{center}
\end{figure}

\begin{table*}
\caption{Sample of 62 ETGs. 
In the left panel 28 ETGs out of the 32
ETGs of the NICMOS sample (Saracco et al. 2009) are listed. 
Four galaxies have not been considered in the present analysis because of 
the very poor fitting to their profile on the HST-NIC3 images (0.2 arcsec/pix).
For each galaxy of the two sub-samples (NICMOS and ACS) we report the total
apparent magnitude (F160W and F850LP) measured on the HST images used to 
perform the profile fitting, the total magnitude of the best fitting
profile derived by \texttt{galfit} (F160W$_{fit}$ and F850LP$_{fit}$),
the effective radius R$_e$, the degree of compactness $C=$R$_e$/R$_{e,z=0}$
(see \S 3), the age and the stellar mass provided by the SED fitting (see \S 2).
The errors on the measured magnitude F160W are in the range
0.04-0.1 mag while those on the F850LP are in the range 0.01-0.05 mag.
The typical error on the best fitting profile magnitude is 0.02 mag.  
}
\centerline{
\begin{tabular}{lcccccccclccccccc}
\hline
\hline
 & & & NICMOS & sample& & & & \vline & &  & & ACS& sample & & & \\
 \hline 
  ID  & $z_{spec}$ &F160W& F160W$_{fit}$& R$_e$ & $C$ & Age & log$(\mathcal{M}_*)$ &\vline  & ID  & $z_{spec}$ &F850LP& F850LP$_{fit}$& R$_e$ & $C$ & Age & log$(\mathcal{M}_*)$\\
          &    &[mag] &[mag]  & [kpc] &  &[Gyr] & [M$_{\odot}$] &\vline &    &  & [mag] &[mag]  & [kpc] & & [Gyr] & [M$_{\odot}$] \\
  \hline
S2\_109 & 1.22 &17.75  &17.47	 & 4.4$\pm$0.2 & 0.34 & 3.5  &11.88  &\vline & 01 & 1.921 &  23.96  &   23.84  &  0.5$\pm$0.1 & 0.20  &  1.0 &   10.56 \\
S7\_254 & 1.22 &19.58  &19.46	 & 2.3$\pm$0.2 & 0.26 & 4.5  &11.59  &\vline & 02 & 1.609 &  23.52  &   23.31  &  1.1$\pm$0.1 & 0.25  &  3.2 &   10.99 \\
S2\_357 & 1.34 &19.04  &18.72	 & 2.8$\pm$0.2 & 0.28 & 4.2  &11.69  &\vline & 03 & 1.609 &  23.03  &   22.75  &  1.6$\pm$0.1 & 0.27  &  3.5 &   11.17 \\
S2\_389 & 1.40 &20.08  &19.79	 & 2.1$\pm$0.3 & 0.36 & 3.5  &11.26  &\vline & 04 & 1.610 &  23.53  &   23.33  &  0.8$\pm$0.1 & 0.29  &  1.0 &   10.57 \\
S2\_511 & 1.40 &19.37  &19.15	 & 2.1$\pm$0.2 & 0.54 & 1.0  &10.95  &\vline & 05 & 1.123 &  21.39  &   21.08  &  1.4$\pm$0.1 & 0.36  &  2.4 &   10.81 \\
S2\_142 & 1.43 &19.00  &18.65	 & 3.1$\pm$0.2 & 0.38 & 3.5  &11.61  &\vline & 06 & 1.032 &  22.27  &   22.08  &  0.4$\pm$0.1 & 0.32  &  1.0 &   10.00 \\
S7\_45\ & 1.45 &18.59  &18.83	 & 4.7$\pm$0.3 & 0.77 & 1.0  &11.30  &\vline & 07 & 1.612 &  24.22  &   23.91  &  2.1$\pm$0.3 & 0.42  &  3.0 &   10.99 \\
S2\_633 & 1.45 &19.32  &19.00	 & 2.6$\pm$0.2 & 0.37 & 2.6  &11.40  &\vline & 08 & 1.610 &  23.78  &   23.67  &  0.8$\pm$0.1 & 0.39  &  0.9 &   10.40 \\
S2\_443 & 1.70 &19.69  &19.44	 & 3.4$\pm$0.3 & 0.41 & 3.2  &11.53  &\vline & 09 & 1.044 &  21.32  &   20.98  &  1.4$\pm$0.1 & 0.46  &  2.0 &   10.62 \\
S2\_527 & 1.35 &19.80  &19.50	 & 1.7$\pm$0.3 & 0.39 & 2.3  &11.04  &\vline & 10 & 1.215 &  21.67  &   21.55  &  1.2$\pm$0.1 & 0.47  &  1.0 &   10.60 \\
S-5592  & 1.623&20.40  &20.30	 & 1.4$\pm$0.3 & 0.30 & 0.9  &10.48  &\vline & 11 & 1.614 &  23.30  &   22.92  &  2.0$\pm$0.2 & 0.54  &  1.7 &   10.75 \\
S-5869  & 1.510&19.64  &19.53	 & 2.8$\pm$0.3 & 0.35 & 1.2  &10.60  &\vline & 12 & 0.964 &  20.05  &   19.98  &  3.6$\pm$0.1 & 0.56  &  3.0 &   11.31 \\
S-6072  & 1.576&21.06  &20.96	 & 1.4$\pm$0.3 & 0.45 & 1.4  &10.48  &\vline & 13 & 1.910 &  23.33  &   22.97  &  2.9$\pm$0.2 & 0.57  &  0.9 &   11.00 \\
S-8025  & 1.397&19.94  &19.87	 & 2.4$\pm$0.3 & 0.52 & 3.7  &10.70  &\vline & 14 & 1.096 &  20.32  &   20.20  &  4.2$\pm$0.4 & 0.60  &  2.7 &   11.36 \\
S-8895  & 1.646&19.44  &19.20	 & 3.9$\pm$0.3 & 0.49 & 0.8  &10.85  &\vline & 15 & 1.096 &  21.46  &   21.46  &  2.4$\pm$0.3 & 0.61  &  3.2 &   10.96 \\
S-4367  & 1.725&20.75  &20.61	 & 2.5$\pm$0.3 & 0.88 & 0.9  &10.60  &\vline & 16 & 1.604 &  23.87  &   23.72  &  1.2$\pm$0.1 & 0.58  &  1.1 &   10.38 \\
S-5005  & 1.845&20.59  &20.46	 & 2.1$\pm$0.3 & 0.67 & 0.9  &10.60  &\vline & 17 & 1.297 &  22.35  &   22.27  &  2.0$\pm$0.2 & 0.64  &  2.3 &   10.74 \\
S-7543  & 1.801&19.70  &19.64	 & 3.3$\pm$0.3 & 0.79 & 1.0  &10.95  &\vline & 18 & 1.097 &  22.44  &   22.15  &  1.2$\pm$0.1 & 0.61  &  1.9 &   10.33 \\
S-0189  & 1.490&19.27  &19.19	 & 3.2$\pm$0.3 & 0.40 & 3.5  &11.26  &\vline & 19 & 1.125 &  21.78  &   21.48  &  2.1$\pm$0.1 & 0.66  &  2.1 &   10.66 \\
S-1983  & 1.488&20.02  &20.03	 & 1.5$\pm$0.3 & 0.32 & 3.7  &11.00  &\vline & 20 & 1.039 &  21.28  &   20.97  &  2.4$\pm$0.1 & 0.68  &  2.5 &   10.74 \\
C\_237  & 1.271&20.38  &20.14	 & 3.0$\pm$0.6 & 1.40 & 3.5  &10.48  &\vline & 21 & 1.022 &  20.84  &   20.41  &  3.8$\pm$0.1 & 0.74  &  2.7 &   11.00 \\
C\_65   & 1.263&18.71  &18.85	 & 3.3$\pm$0.3 & 0.52 & 4.2  &11.32  &\vline & 22 & 1.019 &  22.92  &   22.72  &  0.6$\pm$0.1 & 0.65  &  1.1 &	  9.77 \\
C\_142  & 1.277&19.67  &19.63	 & 1.6$\pm$0.4 & 0.38 & 4.2  &11.00  &\vline & 23 & 1.089 &  20.67  &   20.34  &  3.2$\pm$0.1 & 0.78  &  0.8 &   10.86 \\
C\_135  & 1.276&19.37  &19.33	 & 4.7$\pm$0.4 & 1.19 & 4.3  &10.95  &\vline & 24 & 0.980 &  20.17  &   19.83  &  3.7$\pm$0.1 & 0.80  &  1.6 &   10.94 \\
H\_1031 &1.015 &19.57  &19.37	 & 2.1$\pm$0.2 & 1.24 & 1.1  &10.30  &\vline & 25 & 0.964 &  20.12  &   19.69  &  5.7$\pm$0.1 & 0.95  &  2.6 &   11.11 \\
H\_1523 &1.050 &18.00  &17.67	 & 4.8$\pm$0.8 & 0.79 & 2.0  &11.32  &\vline & 26 & 1.415 &  23.45  &   23.26  &  2.1$\pm$0.2 & 0.87  &  1.1 &   10.48 \\
H\_ 731 &1.755 &20.24  &20.20	 & 4.6$\pm$0.8 & 1.84 & 1.4  &10.60  &\vline & 27 & 1.221 &  22.92  &   22.55  &  2.2$\pm$0.2 & 0.91  &  2.4 &   10.43 \\
W091    & 1.55 &19.64  &19.77	 & 1.6$\pm$0.2 & 0.51 & 2.4  &10.78  &\vline & 28 & 1.041 &  21.59  &   21.19  &  2.8$\pm$0.1 & 0.97  &  1.9 &   10.55 \\
---     & ---  & ---   & ---	 &     ---     & ---  & ---  & ---   &\vline & 29 & 1.188 &  22.78  &	22.45  &  2.4$\pm$0.2 & 1.01  &  2.5 &   10.44 \\
---     & ---  & ---   & ---	 &     ---     & ---  & ---  & ---   &\vline & 30 & 1.222 &  22.27  &	22.06  &  2.4$\pm$0.1 & 1.10  &  1.1 &   10.42 \\
---     & ---  & ---   & ---	 &     ---     & ---  & ---  & ---   &\vline & 31 & 1.135 &  20.53  &	20.28  &  9.3$\pm$0.9 & 1.30  &  2.6 &   11.32 \\
---     & ---  & ---   & ---	 &     ---     & ---  & ---  & ---   &\vline & 32 & 1.170 &  22.31  &   21.87  &  3.5$\pm$0.2 & 1.34  &  2.4 &   10.46 \\
---     & ---  & ---   & ---	 &     ---     & ---  & ---  & ---   &\vline & 33 & 1.330 &  21.95  &	21.31  &  7.9$\pm$0.7 & 1.48  &  1.8 &   10.93 \\
---     & ---  & ---   & ---	 &     ---     & ---  & ---  & ---   &\vline & 34 & 0.984 &  21.29  &	20.73  &  4.4$\pm$0.4 & 1.43  &  1.7 &   10.54 \\
\hline									 
\hline									 
\end{tabular}								 
}									 
\end{table*}

\section{Probing the build-up of normal and compact early-type galaxies}
Fig. 1 shows the relation between the effective radius R$_e$ [kpc]
and the stellar mass $\mathcal{M}_*$ [M$_\odot$] of the 62
ETGs considered in the present work (filled symbols).
{ The different colors used in the upper panel mark the different 
redshift interval which the galaxies belong to: 
blue corresponds to $0.9<z_1<1.2$, cyan to $1.2<z_2<1.5$ and magenta
to  $1.5<z_3<2$.}
The size-mass (SM) relation found by Shen et al. (2003) for the 
local population of ETGs is also shown (solid line). 
The fraction of normal ETGs, which we define as those lying within one sigma 
from the local relation { (taking also into account the errors on size 
measurements)}, is $\sim50$ per cent (29 out of 62 ETGs; 
filled triangles), slightly smaller than the 
fraction (62 per cent) derived from the complete ACS sample 
(Saracco et al. 2010).
{ We are not inclined to ascribe the higher fraction of compact ETGs in the
NICMOS sample to a systematic difference of the size of galaxies when observed
at different wavelengths since no evidence in favour of this effect
is emerging (e.g. McGrath et al. 2008; Cassata et al. 2010).}
We are rather inclined to ascribe this difference to the fact that the NICMOS 
sample collects ETGs pre-selected on the basis of their red colors 
(e.g. R-K$>5$), i.e. their old age.
Since old ETGs are smaller for fixed mass and more massive for fixed radius 
(e.g. Bernardi et al. 2008; Valentinuzzi et al 2010)
NICMOS sample is consequently biased toward compact ETGs.
\begin{figure}
\begin{center}
\includegraphics[width=8.6cm]{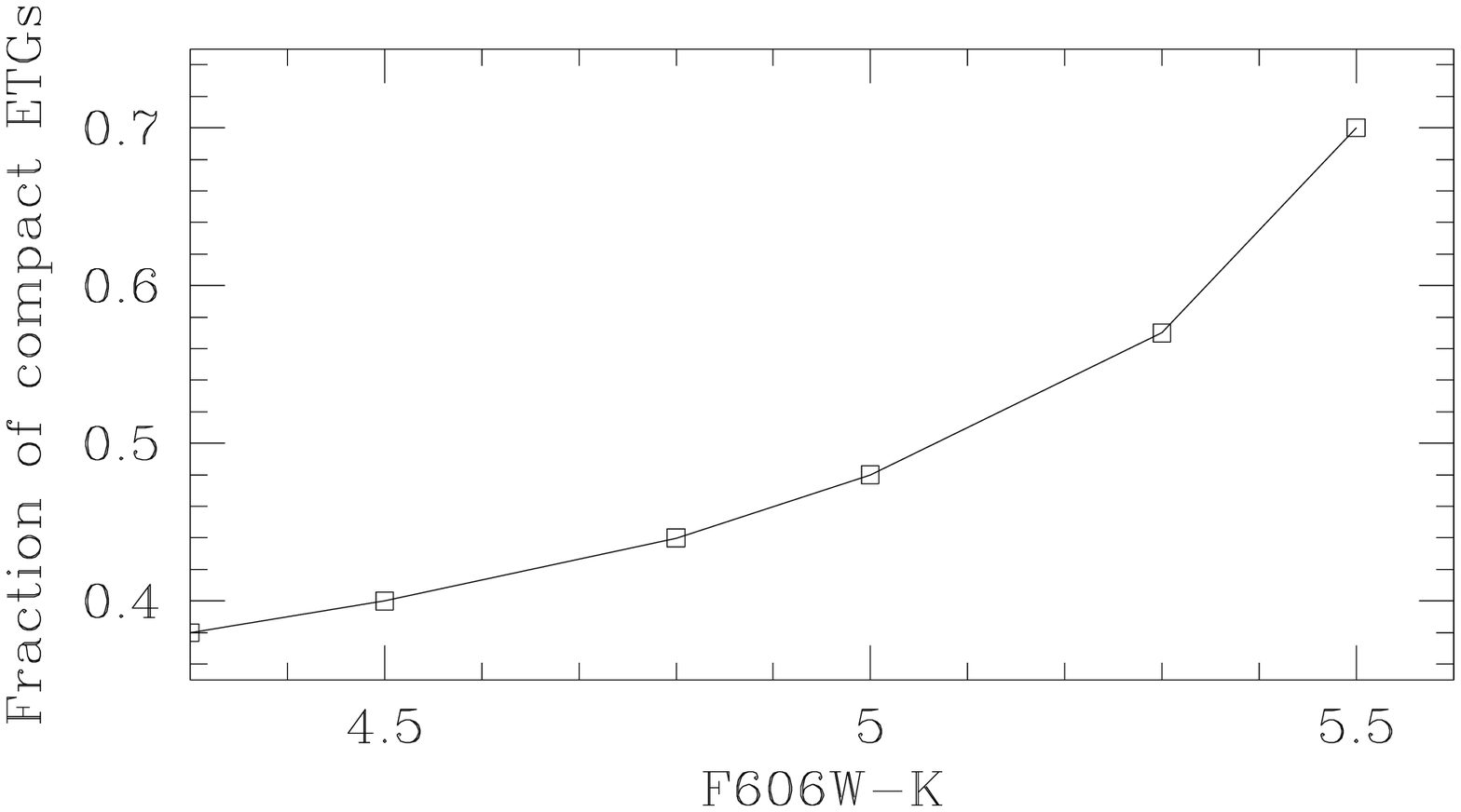} 
\caption{Fraction of compact ETGs (falling below one sigma from the
local SM relation) in the complete ACS sample as a function 
of color F606W-K. It is evident the increasing fraction of compact ETGs
at increasing red color.    
}
\end{center}
\end{figure}
\begin{figure}
\begin{center}
\includegraphics[width=8.6cm]{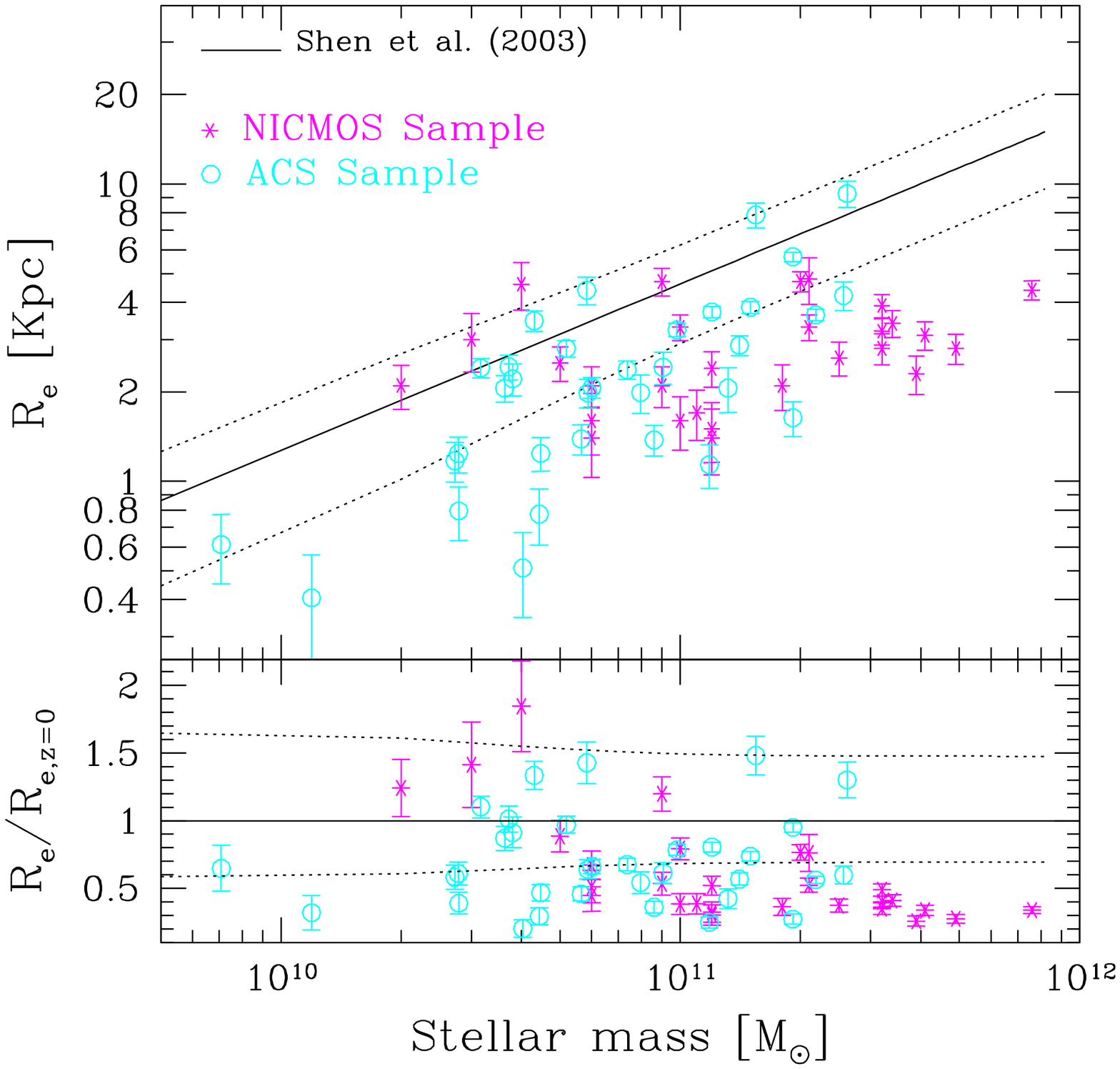} 
\vskip -0.5truecm
\caption{Same as Fig. 1.  In this case the effective radius R$_e$ 
(upper panel) and the compactness (lower panel) of galaxies are plotted 
versus their stellar mass using different symbols according to the
sample they belong to: starred (magenta) symbols mark the ETGs
of the NICMOS sample while open (cyan) circles the ETGs of the ACS sample. 
It is evident the larger fraction of compact ETGs in 
the NICMOS sample, particularly at  $\mathcal{M}_*\gae 10^{11}$ 
where all the ETGs are compact.}
\end{center}
\end{figure}
We verified this by using the complete ACS sample, unbiased
with respect to any color selection.
We have selected galaxies from this sample at different F606W-K colors
(close to R-K color) and we have count the fraction of compact
ETGs, which we define as those falling more than one sigma below the local 
size-mass relation, at different color cuts. 
The result is shown in Fig. 2 where the fraction of compact ETGs is shown 
as a function of F606W-K color cuts. 
It is evident that the fraction of compact ETGs increases systematically
toward redder color cuts showing the strong selection effect affecting the
NICMOS sample and all those ETG samples constructed using this kind of color 
pre-selection.
{ For clarity, in Fig. 3 the same relations shown in Fig. 1 are presented  
using different symbols (and colors) for the galaxies belonging to the two 
sub-samples: starred (magenta) symbols  represent the ETGs of the NICMOS 
sample while open (cyan) circles those belonging to the ACS sample.
It is evident the different distribution of the NICMOS sample
with respect to the ACS sample in the size-mass plane.
Most importantly, it is evident the larger fraction of compact ETGs in 
the NICMOS sample, particularly at large stellar masses 
($\mathcal{M}_*\gae 10^{11}$) where all the NICMOS ETGs are compact.}
This is the reason why, in Fig. 1 at masses $\mathcal{M}_*>3\times 10^{11}$ 
M$_\odot$, a range populated only by galaxies belonging to the NICMOS sample
(see Fig. 3) there are only compact ETGs.  
In spite of this, Fig. 1 (lower panel) shows that  the compactness of a galaxy, 
defined as the ratio between its effective radius R$_e$ and the effective radius 
R$_{e,z=0}$ of an equal mass galaxy at $z=0$ derived from the local SM 
relation, does not show any trend with mass.
This evidence conflicts with the hypothesis of a mass-dependent size 
evolution of ETGs, hypothesis suggested by many authors.
For instance, it has been suggested that only high-mass ETGs undergo  
size evolution (e.g. Newman et al. 2010) or that they undergo the
most rapid evolution (e.g. Ryan et al. 2010) being them on average smaller 
than their local counterparts.
At the same time, other authors have suggested at odd with this 
that only low-mass ETGs  undergo size evolution to match the apparent 
lack of compact low-mass ETGs in the local Universe 
(e.g. van der Wel et al. 2009).
Actually, we find that normal and compact ETGs co-exist for large 
intervals of effective radius (0.4 kpc $<R_e<$ 8 kpc) and of the stellar 
mass ($10^{10}$M$_\odot<\mathcal{M}_*\lae10^{12}$M$_\odot$)
as already found on a smaller but complete sample of ETGs 
(Saracco et al. 2010).

In Fig. 4 (left-hand panels) the compactness R$_e$/R$_{e,z=0}$
(upper panel), the stellar mass $\mathcal{M}_*$ (middle panel) and the 
effective stellar mass density 
$\rho_e=0.5\mathcal{M}_*/({4\over3}\pi R_e^3)$ 
(we assumed that M/L is radially constant; lower panel)
 for the 62 ETGs of our sample are shown as a 
function of their redshift.
It is evident the co-existence of normal-size ETGs  
(R$_e\simeq$R$_{e,z=0}$ and $10^7<\rho_e<10^9$ M$_\odot$ kpc$^{-3}$) 
with ETGs of the same mass but with 
R$_e\simeq[0.5-0.2]\times$R$_{e,z=0}$, that is 10-100 
times denser ($\rho_e>10^9$ M$_\odot$ kpc$^{-3}$).
The origin of this extremely large spread in the properties 
of ETGs at $0.9<z<2$  necessarily originates before $z\sim2$, during 
their assembly.
Therefore, we searched for differences in their past history  
to sketch a possible scenario.

\begin{figure*}
\begin{center}
\includegraphics[width=18.cm]{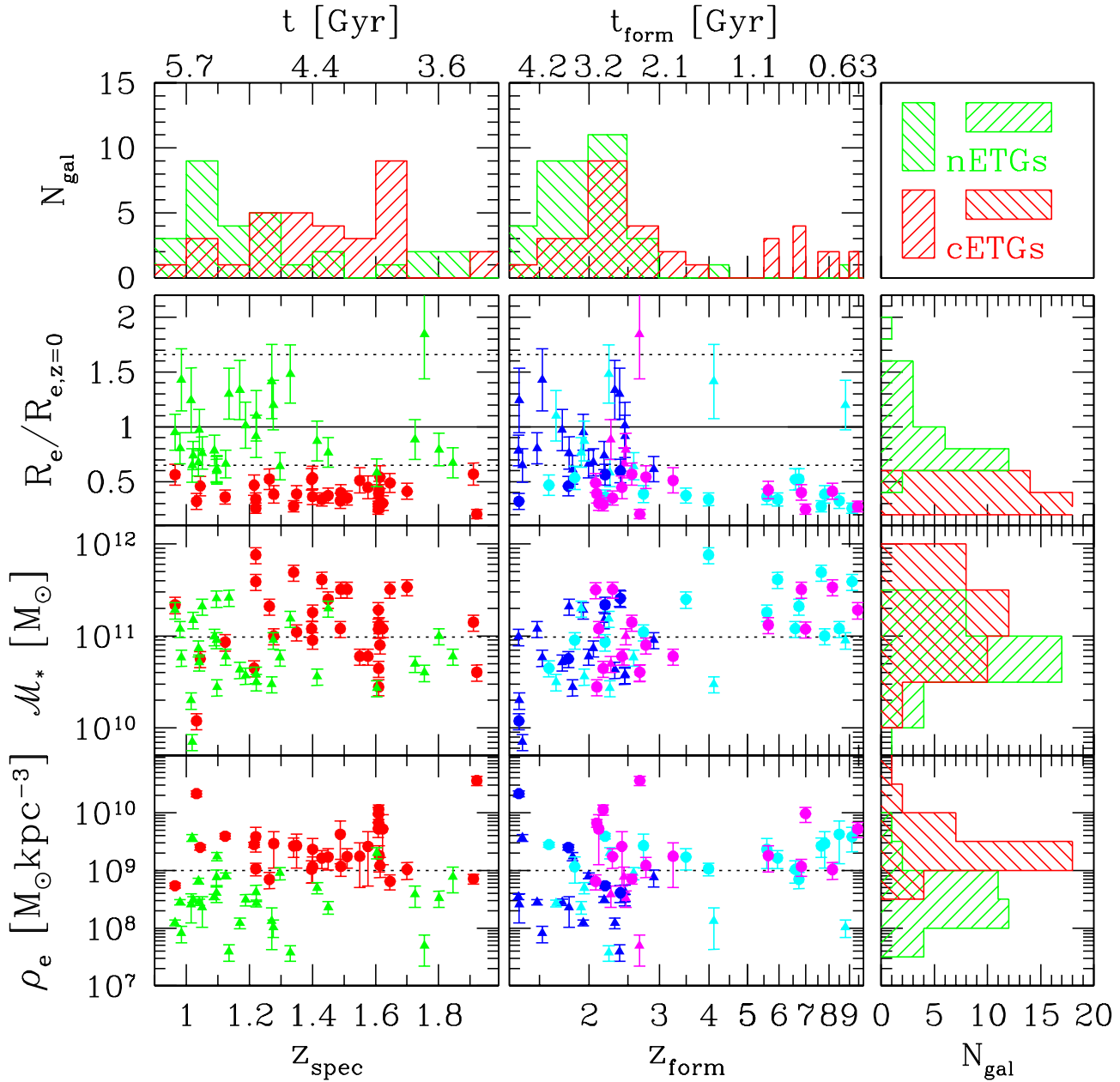} 
\vskip -1truecm
\caption{The compactness R$_e$/R$_{e,z=0}$ (upper panels), the stellar mass 
$\mathcal{M}_*$ (middle panels) and the effective stellar mass density 
 $\rho_e=0.5\mathcal{M}_*/({4\over3}\pi R_e^3)$ (lower panels) 
of a galaxy are shown as a function of redshift $z_{spec}$ 
(left-hand panels) and of the formation redshift $z_{form}$ (right-hand panels) 
for the whole sample of 62 ETGs at $0.9<z_{spec}<2$. 
Triangles and circles mark normal and compact ETGs respectively
in all the panels.
{ 
The histograms at the top of the figure represent the distributions of redshift 
(left-hand) and of formation redshift (right-hand) for normal  
ETGs (nETGs, green shaded) and for compact ETGs (cETGs, red shaded). 
On the upper x-axis the age of the universe at the different redshift 
$z_{spec}$ and at the different formation redshift $z_{form}$ 
is shown.
The histograms at the right-hand side of the figure represent
the distributions of the compactness (upper panel), of the stellar mass
(middle panel) and of the effective stellar mass density (lower panel) 
for normal and compact ETGs.
In the left-hand panels normal and compact are identified by green and 
red colors respectively.
In the right-hand panels, showing quantities as a function of $z_{form}$, 
galaxies are marked with different colors (blue, cyan and magenta)
according to the redshift range they belong to ($0.9<z_1<1.2$, $1.2<z_2<1.5$
and  $1.5<z_3<2$ respectively), as in Fig. 1. 
The solid line in the compactness vs $z_{spec}$ and $z_{form}$ panels 
represents the relation $R_e/R_{e,z=0}=1$ while the dotted 
lines are the 1 sigma dispersion of the local SM relation measured 
at $10^{11}$ M$\odot$.
In the middle and in the lower panels the dotted lines at 
$\mathcal{M}_*=10^{11}$ M$_\odot$ and at $\rho_e=10^9$ M$_\odot$ kpc$^{-3}$
mark the median values.}
}
\end{center}
\end{figure*}

\subsection{Constraining the formation of the stellar mass} 
There is not  direct gauge of how and when the stellar mass has been 
assembled to form and to shape an early-type galaxy. 
However, the age of the stellar population fixes at least the 
epoch at which the stellar mass formed even if, according to the present 
models of galaxy formation, mass assembly and epoch of star formation are 
not necessarily concurrent (e.g. De Lucia et al. 2006). 
We derived for each galaxy the formation redshift 
$z_{form}=z(t_{form})$, 
where $t_{form}=age_{Univ}(z_{spec})-age_{star}(z_{spec})$ is the epoch at 
which the bulk of the stellar mass was already formed,
$age_{Univ}(z_{spec})$ is the age of the universe at the redshift of the galaxy
and $age_{star}(z_{spec})$ is the age of the stellar population 
provided by the best fitting model to the observed
SED of the galaxy at its redshift.
In the right-hand panels of Fig. 4 the compactness, the stellar mass
and the effective stellar mass density  are shown as a 
function of their  formation redshift $z_{form}$.
The different colors blue, cyan and magenta mark the different
spectroscopic redshift ranges 0.9-1.2, 1.2-1.5 and 1.5-2 
which the galaxies belong to.
{ When the redshift (left-hand panels) is considered normal and
compact do not show any convincing trend with $z_{spec}$.}
Their redshift distributions, shown at the 
top of Fig. 4, peak at different redshift, higher for compact ETGs and
lower for normal ETGs even if they extend over the same redshift 
range.  
The peak at $z_{spec}\simeq1.6$ of compact ETGs is due to the high-density 
sheet-like structure present in the GOODS-South field  (Kurk et al. 2009)
containing 5 compact ETGs.
The two distributions are different at $\sim$95 per cent confidence level being
 P$_{KS}\simeq5\times10^{-2}$ the probability of the two samples 
 Kolmogorov-Smirnov (KS) test.
{ The increasing fraction of compact ETGs with redshift
(from 29 per cent (6/21) at $z_{spec}=0.9-1.2$ to 64 per cent (14/22) at
$z_{spec}=1.2-1.5$)  
also visible in Fig. 1 when the whole sample of 62 ETGs is considered,
is actually due to the high number of compact ETGs present in the NICMOS sample.
Indeed, if we consider the complete ACS sample, the fraction of
compact ETGs in the same redshift ranges are 21 per cent (4/19) 
and 17 per cent (1/6) respectively.
However, we have to note that in the highest redshift bin $z_{spec}=1.5-2$ 
the fraction of compact ETGs increases to 50 per cent (2/4, not considering
the overdensity) even if the very low statistics does not make significant 
this difference.} 
When $z_{form}$ is considered almost all (27 out of 29) the 
normal ETGs are segregated at $z_{form}<3$, while compact ETGs are 
distributed over a much wider interval 
with half of them (16 out of 33) at $z_{form}>3$ and 40 per cent, 13 out of 33,
at $z_{form}>5$ (right-hand diagrams). 
The distributions of the formation redshift of compact 
and normal ETGs (histogram at the top of Fig. 4) differ at 99.95 per cent 
confidence level (P$_{KS}\simeq5\times10^{-4}$).
The difference between the two $z_{form}$ distributions is represented by 
the large number of compact ETGs with $z_{form}>5$ (and $z_{spec}<1.5$), 
an interval actually populated only by them.
Also, Fig. 4 (middle-right panel) shows that they are massive 
($\mathcal{M_*}>10^{11}$ M$_\odot$) and, consequently, highly dense 
($\rho_e\gae 10^9$ M$_\odot$ kpc$^{-3}$; lower-right panel). 
On the contrary, at $z_{form}<3$ there are both normal and 
compact galaxies independently of stellar mass and mass density.
{ In this connection, we note that the distributions
of the stellar mass of normal and compact ETGs shown at the right-hand
side of Fig. 4 differ at 99 per cent confidence level 
(P$_{KS}\simeq8\times10^{-3}$) due to the high-mass tail  
($\mathcal{M}_*>3\times 10^{11}$ M$_\odot$) of 
compact ETGs.
We point out that the distributions of the compactness R$_e$/R$_{e,z=0}$ and
of the effective stellar mass density $\rho_e$ of normal and compact ETGs  
differ by definition of normal and compact galaxy. }
Strictly speaking, Fig. 4 shows that earlier stars, those 
characterized by $z_{form}>5$ 
are preferentially assembled in compact, more massive and hence denser ETGs
while stars later formed ($z_{form}<3$), or  possibly resulting 
from subsequent episodes of star formation, are assembled
both in compact and in normal ETGs independently of their mass.
It is worth noting that this behaviour only partially agrees with the 
"downsizing" pattern  (Cowie et al.1996; Gavazzi et al. 2002) where the 
higher the mass the older the stellar population and hence the higher the 
formation redshift.
Indeed, Fig. 4 shows that the older the stellar population the higher the 
mass, but not vice versa.
Indeed, we see ETGs with $z_{form}<3$ with masses as large as those
with $z_{form}>5$.
Thus, it seems that the epoch of formation may  play a role 
in the formation of massive ETGs rather than the mass itself. 
Given the rapid decline ($0.1\le\tau\le0.3$ Gyr) of the SFHs 
($\tau^{-1}e^{-t/\tau}$) of our ETGs it follows that 
the bulk  of the stellar mass is produced within 1 Gyr.
For those compact ETGs with $z_{form}>5$ this has taken place 
$\approx 12-12.5$ Gyr ago, while for those with $z_{form}<3$ 
at later times or possibly through subsequent episodes.
Actually, these behaviours are qualitatively similar to those  
found in local ETGs. 
For instance, Thomas et al. (2005; 2010) studying a local sample of early-type
galaxies, showed that the peak of the star formation efficiency occurred 
12 Gyr ago for the most massive ETGs and that the higher the 
mass the shorter and more efficient the burst.
Gargiulo et al. (2009), studying the foundamental plane (FP) of local ETGs 
find that for a given mass galaxies more compact than expected from
the FP relation have experienced their last burst of star formation at 
earlier epochs.
We tried to put in the context of galaxy formation models our
results on the star formation histories of high-z ETGs in 
order to constrain their possible assembly history.

\subsection{Constraining the assembly of the stellar mass}
In the current paradigm of galaxy formation
the build-up of ETGs follows a hierarchical merging scheme along which 
mergers between sub-units can be dissipative (gas-rich) or 
dissipation-less ("dry").
This scheme has to account for the co-existence at $1<z<2$ of ETGs 
already shaped and grown in mass 
($10^{10}$M$_\odot<\mathcal{M_*}<10^{12}$M$_\odot$),
with very different stellar mass densities ($10^{8}\lae\rho_e\lae 10^{10}$ 
M$_\odot$ kpc$^{-3}$) and with stellar populations apparently formed  at 
different epochs ($1.5-2<z_{form}<10$). 
Given the redshift of our ETGs ($1<z<2$) it follows that the above non 
homogeneity must be realized at $z>2$.
This leads to 3-4 Gyr the time at disposal to realize the above properties. 

Let us first consider compact ETGs at $<z>\simeq1.5$.
They should be a natural consequence of the dissipative spheroid formation.
Indeed, gas-rich merger is the known mechanism able to build compact ETGs on 
condition that the starburst resulting from the central dissipative gas 
collapse produces a large fraction ($\gae50$ per cent) of the stars of the 
remnant
(e.g. Springel \& Hernquist 2005; Khochfar \& Silk 2006a, 2006b; 
Naab et al. 2007, 2009; Ciotti et al. 2007; Hopkins et al. 2008). 
Actually, the higher the fraction of stars formed in the merger the 
smaller the remnant at fixed mass (e.g. Khochfar \& Silk 2006a).
{ A typical time scale of merging  in case of nearly equal mass
merging galaxies is $\tau_{merge}\simeq\tau_{dyn}\simeq1.5$
Gyr (e.g. Boylan-Kolchin et al. 2008), where $\tau_{dyn}$ is the 
dynamical time.
Naab et al. (2007) found that their simulated galaxies start forming
galaxies at $z\simeq3-5$ concurrently with the intense phase of merging
and that after about 1.5-2 Gyr ($z\simeq2.5-2$) they were assembled 
almost 80 per cent of their final stellar mass (see also Sommer-Larsen
and Toft 2010).   
}
Thus, for a compact ETG  the assembly of the stellar mass 
can be considered nearly concurrent with its formation, that is 
$z_{assembly}\simeq z_{form}$.
Stated this, the constrains on the formation of the stellar mass of compact 
ETGs derived above (\S 3.1) suggest two possible interpretations for their assembly.
The first one is that if the $z_{form}$ values are reliable for all the compact
ETGs then, given the formation redshift $2\lae z_{form}<10$ (see Fig. 4), 
it follows that the assembly of compact ETGs has taken place over the range 
$2<z_{assembly}<10$.
This implies that the physical conditions able to produce compact ETGs 
(dissipative gas-rich merger) occur independently of redshift, at least in 
the above redshift range.
According to this interpretation, $\sim40$ per cent of the compact ETGs 
in our sample, those with $z_{form}>5$, assembled at $z_{assembly}\gae5$.  
Perhaps, we should more properly say that the first ETGs assembled 
at $z_{assembly}\gae5$, were compact and more massive than 
$10^{11}$ M$_\odot$ (right-hand panels of Fig. 4). 
The other possibility is that minor bursts of star formation 
occurred later have made some of the compact ETGs looking younger lowering 
the $z_{form}$, that is the low $z_{form}$ values (e.g. $z_{form}<3-5$) are
not reliable.
It is important to point out that in a dissipative merger where a large 
fraction of stars necessarily results from the central starburst ignited
by the merger itself, it is easy to lower the $mean$ age of the resulting 
stellar population and hence the formation redshift hypothesizing 
one or more small starburst episodes occurred later while it is difficult 
to realize the opposite condition.
{ The spectra of our galaxies do not show signs of ongoing star formation
and we do not find any systematic difference in the SED fitting of 
$z_{form}>5$ and $z_{form}<3$ compact ETGs.
We also performed the fitting by neglecting in turn the UV rest-frame 
data of the SED dominated by the youngest stars and the near-IR rest-frame 
data dominated by the oldest stars without obtaining differences in the 
best fitting templates.
However, this is not sufficient to exclude the small starbursts hypothesis 
since even a single burst adding a few per cent of young stars can
affect the mean age of the stellar population of a galaxy
(e.g. Longhetti et al. 2005; Serra \& Trager 2007; Thomas et al. 2010).}
Thus, in the hypothesis of one or more small starburst(s) occurred later, 
not only those with $z_{form}>5$ but also 
the other compact ETGs could have been assembled at $z_{assembly}\gae5$
according to the reasonable hypothesis that the occurrence of dissipative 
mergers is dependent on redshift, much more probable at high-z given
the larger fraction of gas at disposal (e.g. de Lucia et al. 2006).
To summarize, our results suggest that in a dissipative merging scheme 
the assembly of compact ETGs has taken place at high-z 
($z_{assembly}>5$) for most of them or for a large fraction (40\%
in the present sample) of them at least.
This latter case requires that dissipative gas-rich merger can occur 
efficiently over a wide redshift range, at least at $z>1$.

Let us now consider normal ETGs at $<z>\simeq1.5$.
We notice that normal ETGs span nearly the same mass range of compact ETGs but 
their stellar population seems to be formed more recently ($z_{form}<3$).
The main hypothesis to probe is whether they are descendants of the
compact/dense ETGs assembled at high-z.
Actually, we investigated the general scheme in which all the ETGs assembled 
at high-z as compact/dense spheroids and a fraction of them grow in size
by adding a low stellar mass density envelope through subsequent gas-poor 
minor mergers at later times (e.g. Hopkins et al. 2009; Naab et al. 2009; 
Bezanson et al. 2009).
Given the constraints on the redshift of assembly of compact ETGs derived 
above ($z_{assembly}>5$) and the redshift range covered by our sample
($0.9<z<2$) it follows that the possible subsequent gas-poor minor 
mergers should occur at $1.5-2\lae z<5$,  that is a fraction of the compact 
spheroids formed at $z>5$ should increase  their size from 2 to 6 times in 
about 2.5 Gyr.
The above fraction corresponds to a number density not lower than 
$(5.5\pm3)\times10^{-5}$ Mpc$^{-3}$, which is the number density of normal 
ETGs observed at $\langle z\rangle=1.5$ with masses $[0.1-4]\times10^{11}$ 
M$_\odot$ (Saracco et al. 2010).
{ It is worth noting that in the case of minor mergers the typical time 
scales are $\tau_{merge}>3-4$ Gyr (e.g. Boylan-Kolchin et al. 2008), 
thus larger than the 2.5 Gyr at disposal.}
Simulations suggest that the effective radius increases with merging as 
R$_e\propto\mathcal{M}^\alpha$ with $0.6<\alpha<1.3$ 
(e.g. Boylan-Kolchin et al. 2006; Khochfar \& Silk 2006a, 2006b; 
Ciotti et al. 2007).
However, recently, it has been shown that the accretion of a low stellar 
mass density envelope through dry minor merging can be even more efficient in 
enlarging the size (Naab et al. 2009).
Thus, for our calculations we considered  the extreme value 
$\alpha=1.5$.
In this extremely favourable case a galaxy should at least double its mass 
to increase its effective radius by a factor 3.
Assuming a maximum mass ratio 1:3 for the merging galaxies involved
in minor mergers, at least 3 minor mergers are needed to double the 
mass of a compact ETG in 2.5 Gyr.
We used the merger rate calculator described by Hopkins et al. (2010b)
to derive the number of minor mergers which a galaxy in 
the mass range $[0.1-4]\times10^{11}$ M$_\odot$ can experience at 
$1.5-2\lae z<5$.
We considered a maximum mass ratio 1:3 and
assumed a maximum gas fraction of merging galaxies of 0.2.
The merger rate per galaxy we derived at $z=4$ is 0.013 Gyr$^{-1}$
and thus, keeping constant this rate between $1.5-2\lae z<5$, 
the number of mergers per galaxy in 2.5 Gyr is just 0.03, two orders
of magnitude lower than the one needed.
Even relaxing the gas-poor hypothesis by increasing the minimum gas fraction 
up to $\sim60$ per cent, we obtain only one merger event in 2.5 Gyr
(a merger rate of $\sim0.4$ Gyr$^{-1}$).
Thus, it seems that the minor merger rate predicted by the models is too low
to account for the size increase of early compact spheroids needed in 
the hypothesis that normal ETGs are their descendants.
Our conclusion agrees with the one of Newman et al. (2010) and with the 
conclusions reached by Tiret et al. (2010) on the basis of different 
arguments.

\begin{figure}
\begin{center}
\includegraphics[width=8.6cm]{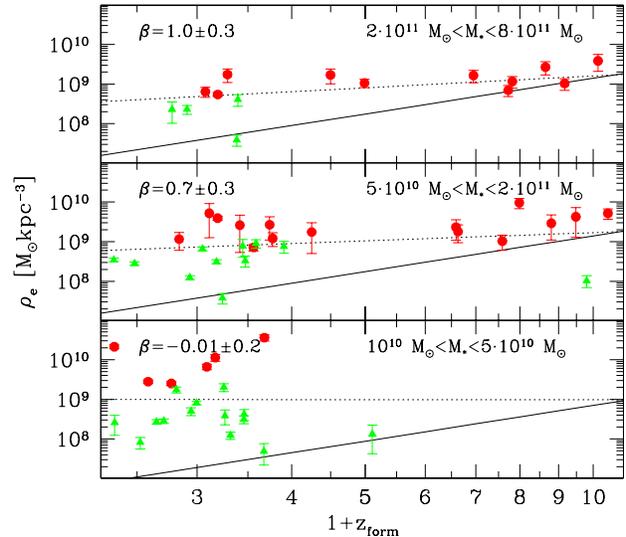} 
\vskip -0.8truecm
\caption{Effective stellar mass density $\rho_e$  
as a function of formation redshift $z_{form}$ for ETGs in different
ranges of stellar masses.
The dotted line represent the best fit to the data obtained by fitting
the relation
$\rho_e(z_{form})\propto (1+z_{form})^\beta$.
The solid line is the relation 
$\rho_e(z_{form})\propto (1+z_{form})^3$
expected in the hypothesis that the higher the redshift of 
assembly the denser the ETG due to the higher density of the 
Universe at earlier epochs. 
The expected relation is normalized to the best fitting value 
at $z_{form}=10$.
We considered the approximation $z_{assembly}\simeq z_{form}$.
}
\end{center}
\end{figure}

Another proposed scheme is that  the higher the redshift of 
assembly the denser the ETG due to the higher density of the 
Universe at earlier epochs.
{ Since the linear size varies as $(1+z)^{-1}$,
in the approximation $z_{assembly}\simeq z_{form}$, 
the effective stellar mass density of a galaxy
of fixed mass assembling at different $z_{assembly}$ would scale 
in this scheme as 
$\rho_e(z_{assembly})\simeq\rho_e(z_{form})\propto (1+z_{form})^3$ 
independently of the mass of the galaxy.
In Fig. 5 the effective stellar mass density $\rho_e$ of ETGs in 3 
different ranges of stellar masses is shown as a function of their $z_{form}$.
It is interesting to note that the median effective stellar mass
density is $\langle\rho_e\rangle\simeq10^9$ M$_\odot$ kpc$^{-3}$
in all the three mass ranges.
By fitting the data with a scaling relation 
$\rho_e(z_{form})\propto (1+z_{form})^\beta$ (dotted line)
we obtained $\beta>0$ in the first two ranges of stellar masses considered 
($\beta=1.0\pm0.3$ and  $\beta=0.7\pm0.3$ respectively).
The significance of these mild correlations is about 90 per cent the 
Spearmen rank test probability being $P_{rs}\gae0.08$.  
In Fig. 5  the expected scaling relation 
$\rho_e(z_{form})\propto (1+z_{form})^3$  
normalized to the value of the best fitting relation at $z_{form}=10$
is also shown for comparison (solid line).} 
Thus, the correlation between the effective stellar mass density and
$z_{form}$ is not so evident from our data as expected in the above scenario.
However, we note that $z_{form}$ could not always be a tracer of 
the assembly epoch, as previously stated, and that the statistics is still low.
For these reasons, the lack of the steep scaling relation expected cannot be 
considered a definite evidence against the above picture.

The results shown in Fig. 4 suggest, at the same time, more naturally 
the following scheme of formation and assembly.
Assuming that dissipative gas-rich merger is the mechanism of spheroids 
formation, compact ETGs would be a natural consequence of this mechanism
when most of the gas at disposal is burned in the central starburst, 
as discussed above. 
To this end, the gas involved in the merger has to be sufficiently 
cold to collapse toward the center and then ignite the main starburst
producing the compact remnant.
However, we could suppose that the gas in some of the progenitors is not
so cold and homogeneous to allow the rapid central collapse described above 
and that, consequently, the resulting starburst is not short and intense 
but much longer and possibly composed of many subsequents starbursts.
The rate at which the starburst(s) would be ignited or stoked could be 
modulated by the cooling time of the different gas clouds composing the gas 
reservoir and by their orbital and dynamical parameters.
In this case the remnant would not be compact and
the mean age of the resulting stellar population would be much younger than 
the one produced in a single and short burst.
This qualitative scenario would explain the co-existence of normal and compact 
ETGs observed at $\langle z\rangle\simeq1.5$ in spite of the same stellar 
mass, the lack of normal ETGs with high $z_{form}$ and the absence of any 
correlation between compactness, stellar mass and formation redshift.

The study of the spatial distribution of the stellar component of ETGs 
at $z>1$ can provide fundamental information on their past star formation 
and assembly histories.
{ The different way in which the stellar population is assembled
may produce indeed different colour and light profiles.
Such differences should be more pronounced when the stellar populations are
young ($<3-4$ Gyr) hence in the high redshift ETGs. 
For instance, if all the ETGs assemble at high-z ($z_{assembly}>4-5$)
as compact/dense spheroids and a fraction of them grow in size
by adding a low stellar mass density envelope through dry minor mergers
(e.g. Hopkins et al. 2009; Naab et al. 2009) we should observe
compact ETGs characterized by a coeval stellar population centrally peaked
and normal ETGs characterized by a dense core of old stars and an envelope
with flatter shape composed of younger stars.
On the contrary, in the scenario we have proposed above where the cooling of 
the gas modulates the compactness and the mean age of the stellar 
population of galaxies compact and normal ETGs should be characterized
by similar surface brightness profile shape but different colour profiles.
Thus, compact and normal ETGs should be characterized by
different surface brightness  and/or color profiles according to the different
spatial distribution and density of the stellar populations accreted at
different epochs and through different processes.} 
In this regard, we believe that the study of the color gradient  
of high-z ETGs (Gargiulo et al. 2010) 
may represent a powerful probe of the early phases of ETGs formation.

\section{The SFR and the number density of compact spheroids in the very early
Universe}
On the basis of the results discussed above and of the data
we have at hand we have tried to put constraints on the number density of 
compact spheroids assembled in the very early Universe and on the 
resulting contribution to the SFR density. 
The compact ETGs with $z_{form}>5$ have stellar masses
$\mathcal{M}_*=[1-5]\times10^{11}$ M$_\odot$ (Fig. 4, middle-right panel).
Considering a typical stellar mass of about 
$1-2\times10^{11}$ M$_\odot$, according to the gas-rich merger scheme, 
at least 50 per cent of this mass ($\gae6\times10^{10}$ M$_\odot$) should 
form in about 1 Gyr during the merging at $z_{assembly}>5$,
as derived above. 
This implies a mean star formation rate associated to the compact 
remnant $\langle SFR\rangle\simeq60$ M$_\odot$ yr$^{-1}$ at $z>5$.
The two progenitors, with masses (gas+stars) $\sim6\times10^{10}$ M$_\odot$ 
each, cannot have already formed more than $3\times10^{10}$ M$_\odot$ 
(50 per cent of mass) of stars each at the epoch of the merging.
Since they formed these stellar masses before the merging, that is in about 
0.5 Gyr, the required mean SFR is  $\sim60$ M$_\odot$ yr$^{-1}$ also
in this case.
These (relatively low) values agree with those derived for the star-forming 
galaxies observed at $z>5$
(e.g. Hickey et al. 2010; Wilkins et al. 2010) and with the detection  
of massive galaxies ($>10^{10}$ M$_\odot$) at $z>6$ with age 200-700Myr and 
SFR$\sim30$ M$_\odot$ yr$^{-1}$ (e.g. Eyles et al. 2005; 2007).
The possible (expected) intense phase of star formation (of some hundreds 
of M$_\odot$/yr or more) experienced during the dissipative merger would 
last for very short times.  
Indeed, in an exponentially declining star formation history with e-folding
time $\tau\simeq0.1$ Gyr, the star formation rate would drop by a factor 
$\sim3$ in  $0.1$ Gyr and almost by a factor 10 in 0.2 Gyr.
Since 1 Gyr is needed to form most of the stars (see Sec. 3.1), 
the frequency with 
which we would observe a dissipative merger during the intense star
formation phase would be less than 1:10.
This agrees with the apparent lack of strong star-bursting galaxies
at very high-z and with the  SFRs derived for
high-mass spheroidal galaxies observed at $z>2$
(see also Cava et al. 2010 for very recent results). 

We also tried to derive a lower limit to the number 
density of early compact spheroids at $z>5$ and to their contribution 
to the star formation rate density at that redshift.
Out of the 13 compact ETGs with $z_{form}>5$, 3 belong to the ACS 
sample (Saracco et. al 2010) on the GOODS-South field (143 arcmin$^2$) 
and 5 to the sample of massive ETGs selected on the S2F1 field 
(150 arcmin$^2$) (Saracco et al. 2005; Longhetti et al. 2007).
{ The co-moving volumes subtended by these two fields in the
redshift range $4<z<9$, that is within $\sim1$ Gyr at $z\simeq5$} 
are $1.51\times10^6$ Mpc$^{3}$ and $1.64\times10^6$ Mpc$^{3}$ respectively.
Thus, the expected number densities of compact spheroids at $z>5$ derived by
these two small samples are 
$n=2\times10^{-6}$ Mpc$^{-3}$ and $n=3\times10^{-6}$ Mpc$^{-3}$, from
two times lower than the number density ($n=4-5\times10^{-6}$ Mpc$^{-3}$)
of compact ETGs more massive than $10^{11}$ M$_\odot$ observed at 
$\langle z\rangle\simeq1.5$ (Saracco et al. 2010) and at $z=0$
(Valentinuzzi et al. 2010a) 
{ to an order of magnitude lower than the number density
of compact quiescent galaxies expected at $z>2$ (e.g. Wuyts et al. 2009;
Bezanson et al. 2009)}.
The contribution to the co-moving stellar mass density of these early 
compact spheroids is $\sim6\times10^5$ M$_\odot$ Mpc$^{-3}$, to be compared 
with 
$2.5\times10^6$ M$_\odot$ Mpc$^{-3}$, the lower limit to the
co-moving stellar mass density at $z\gae6$ (Eyles et al. 2007).
Finally, their contribution to the star formation rate density (SFRD),
averaged over 1 Gyr at $z>5$, is
$SFRD\simeq6\times10^{-4}$ M$_\odot$ yr$^{-1}$ Mpc$^{-3}$,
an order of magnitude lower than the total SFRD density estimated at $z\lae6$
(Bouwens et al. 2006; Stark et al. 2007).
Thus, all these quantities are well within  those 
derived from the observations of the very high redshift galaxy population.
By the way, these latter observations provide supports
in favour of the very early formation of compact spheroids whose contribution
in terms of stellar mass and star formation rate densities can be significant.

\section{Summary and conclusions}
We used a sample of 62 ETGs at $0.9<z_{spec}<2$ to probe the
star formation history and the mass assembly history of early-type galaxies
at $z>2$.
Using the local size-mass relation as reference we confirm the co-existence at 
$\langle z\rangle\simeq1.5$ of a large number of normal ETGs having
R$_e\simeq$R$_{e,z=0}$  ($10^7<\rho_e<10^9$ M$_\odot$ kpc$^{-3}$) 
with compact ETGs having 
R$_e=[0.5-0.2]$R$_{e,z=0}$ ($\rho_e>10^9$ M$_\odot$ kpc$^{-3}$)
in spite of the same stellar mass and redshift.
We do not see evidence of a dependence of the compactness
and of the stellar mass density $\rho_e$ of ETGs on their stellar mass.

We derived for each galaxy the formation redshift $z_{form}$ at
which most of the stellar mass formed.
We find that normal ETGs are all segregated at $z_{form}\lae3$ while 
compact ETGs are distributed over a much wider range, $2<z_{form}<10$,
with a significant fraction of them (13 out of 33) at $z_{form}>5$.
Earlier stars, those characterized by $z_{form}>5$,
are assembled in compact, more massive ($\mathcal{M}_*>10^{11}$
M$_\odot$) and hence denser  ($\rho_e>10^9$ M$_\odot$ kpc$^{-3}$) ETGs 
that is, the older the stellar population the higher the mass of the 
galaxy but not vice versa.
Indeed, we see many ETGs with $z_{form}<3$ and masses as large as those
with $z_{form}>5$.
Thus, it seems that the epoch of formation may play a role in the 
formation of massive ETGs rather than the mass itself.

We have tried to put the above results in the context of hierarchical models 
of galaxy formation.
In a dissipative merging scheme, the known mechanism able to produce compact 
remnants, where a large fraction of the stellar mass is produced concurrently
with the merging, $z_{assembly}\simeq z_{form}$.
Consequently,  compact ETGs would assemble at $2<z_{assembly}<10$ with a 
significant fraction at $z_{assembly}>5$, according to the $z_{form}$ 
values obtained.
This implies that dissipative gas-rich mergers can efficiently occur also 
at low redshift in spite of the fact that it should be more
probable at high-z thank to the larger amount of gas at disposal.
This suggests that the occurrence of dissipative mergers is also
dependent on other parameters besides the gas at disposal. 
The fact that most of the compact high-z ETGs have a local descendant 
belonging to the galaxy cluster population suggests that the environment 
may play a role in tuning or triggering dissipative mergers.

We then probed the general scheme in which normal ETGs at 
$\langle z\rangle\simeq1.5$ are descendants of compact spheroids 
assembled at $z_{assembly}>5$.
These latter should grow in size (from 2 to 6 times) through dry minor 
mergers during the 2.5 Gyr at $1.5-2<z<5$.
Using the merger rate calculator by Hopkins et al. we estimated that the number
of dry (gas fraction $\le0.2$) minor (1:3) mergers expected in a hierarchical
model at $1.5-2<z<5$ is two orders of magnitude lower than
the one needed. 
To reach the number of mergers comparable to the one needed, it must be dropped 
the dissipation-less requirement (i.e. gas fraction 0.6 at least) and relaxed
the minor merger requirements ($>1:3$).
However, in this case the size would no longer grow with mass as fast as through 
dry mergers and consequently more mergers would be needed exceeding
those predicted by models, and so on. 
Thus, the hypothesis that normal ETGs are the descendants 
of dense early spheroids does not find supports in the current models.

Finally, we do not find evidence supporting a dependence of the compactness
of galaxies on their redshift of assembly, a dependence expected in the 
hypothesis that the compactness of a galaxy is due to  the higher density 
of the Universe at earlier epochs.
The correlation between the effective stellar mass density and
$z_{form}$ expected in the above scenario does not emerge from our data.
However, we remind that $z_{form}$ could not always trace  
the assembly and that the statistics is still low.
Hence, the lack of a correlation cannot be considered a definite
evidence against the above picture.

The results we obtained studying the dependence of mass, compactness and stellar
mass density of ETGs on their formation redshift suggest a clear scenario 
of formation and assembly of the stellar mass, assuming that dissipative 
gas-rich merger is the main mechanism of spheroids formation.
Indeed, compact ETGs would be a natural consequence of this mechanism
when most of the gas at disposal is burned in the central starburst,
that is when the gas is  sufficiently cold to collapse toward the center and 
then ignite the main starburst.
However, if the gas of the progenitors was not sufficiently cold and 
homogeneous the resulting starburst would not be short and intense 
but longer and possibly composed of many subsequents episodes.
The cooling time of the gas clouds and their orbital parameters
could modulate the rate at which the starburst(s) are ignited or stoked.
This mechanism would produce a larger and younger remnant than the
one produced in the short and intense central starburst case.
This qualitative scenario would explain the co-existence of normal and 
compact ETGs observed at $\langle z\rangle\simeq1.5$ in spite of the same 
stellar mass, the lack of normal ETGs with high $z_{form}$ and the absence of 
any correlation between compactness, stellar mass and formation redshift.

{What our analysis shows is that the stellar mass of  ETGs 
results from different  formation histories and that also
the way in which the mass has been assembled to form and 
to shape them is not unique but follows different assembly histories.} 
This araises the question why an ETG follows an 
assembly history instead of  another one.
It is obvious to wonder if the environment plays a role in setting
out the formation and the destiny of an ETG.
The fact that most of the compact high-z ETGs have
a local descendant belonging to the galaxy cluster population 
suggests that the environment can play a role in accounting for 
the diversities (see also Rettura et al. 2010).
Fundamental insights on the past star formation and assembly histories
of ETGs can come from the study of the spatial distribution of their stellar
component.
In this regard we believe that color gradients represent a promising 
tool to investigate the past history of high-z ETGs.

\section*{Acknowledgments}
This work is based on observations made with the ESO telescopes at the Paranal 
Observatory and with the NASA/ESA Hubble Space Telescope, obtained from the 
data archive at the Space Telescope Science Institute which is operated by 
the Association of Universities for Research in Astronomy. 
We thank the referee for the constructive comments.
We acknowledge financial contribution from the agreements ASI-INAF 
I/016/07/0 and I/009/10/0. 

\section{References}

\noindent Bernardi, M., Hyde, J. B., Fritz, A., Sheth, R. K., Gebhardt, K., 
Nichol, R. C. A., 2008, MNRAS, 391, 1191

\noindent Bezanson R., van Dokkum P. G., Tal T., Marchesini D., Kriek M.,
Franx M., Coppi P. 2009, ApJ, 697, 1290

\noindent Bouwens R. J., Illingworth G. D., Blakeslee J., Franx M. 2006,
ApJ, 653, 53

\noindent Boylan-Kolchin M., Ma C.-P., Quataert E. 2006, MNRAS, 369, 1089

\noindent Boylan-Kolchin M., Ma C.-P., Quataert E. 2008, MNRAS, 383, 93

\noindent Buitrago F., Trujillo I., Conselice C. J., Bouwens R. J., Dickinson
M., Yan H. 2008, ApJ, 687, L61

\noindent Cappellari M., di Serego Alighieri S., Cimatti A., et al. 2009,
ApJ, 704, L34

\noindent Carrasco E. R., Conselice C. J., Trujillo I. 2010, MNRAS, 405, 2253

\noindent Cassata P., et al. 2010, ApJ, 714, L79

\noindent Cenarro A. J., Trujillo I. 2009, A\&A, 501, 119

\noindent Chabrier G. 2003, PASP, 115, 763

\noindent Cimatti A., et al. 2004, Nat., 430, 184 

\noindent Cimatti A., et al. 2008, A\&A, 482, 21

\noindent Ciotti L., Lanzoni B., Volonteri M. 2007, ApJ, 658, 65

\noindent Cowie L. L., Songaila A., Hu E. M., Cohen J. G. 1996, AJ, 112, 839

\noindent Daddi E., Renzini A., Pirzkal N., et al. 2005, ApJ, 626, 680

\noindent Damjanov I., McCarthy P. J., Abraham R. G., et al. 2009, ApJ, 
695, 101

\noindent De Lucia G., Springel V., White S. D. M., Croton D., 
Kauffmann G. 2006, MNRAS, 366, 499

\noindent Dunlop J., Peacock J., Spinrad H., Dey A., Jimenez R., Stern D.,
Windhorst R. 1996, Nat. 381, 581

\noindent Eyles L. P., Bunker A. J., Stanway E. R., Lacy M., Ellis R. S.,
Doherty M. 2005, MNRAS, 364, 443

\noindent Eyles L. P., Bunker A. J., Ellis R. S.,  Lacy M., Stanway E. R.,
Stark D. P., Chiu K. 2007, MNRAS, 374, 910

\noindent Fukugita M., Hogan C. J., Peebles P. J. E. 1998, ApJ, 503, 518

\noindent Gargiulo A., et al., 2009, MNRAS, 397, 75

\noindent Gargiulo A., Saracco P., Longhetti M. 2010, MNRAS in press 
[arXiv:1011.2427] 

\noindent Gavazzi G., Boselli A., Pedotti P., Gallazzi A., Carrasco L. 2002,
A\&A, 396, 449

\noindent Giavalisco, M., Dickinson, M., Ferguson, H. C., et al. 2004,
ApJ, 600, L103

\noindent Glazebrook K., et al. 2004, Nat. 430, 181

\noindent Hickey S., Bunker A., Jarvis M. J., Chiu K., Bonfield D. 2010,
MNRAS, 404, 212

\noindent Hopkins, P. F., Cox T. J., Hernquist L. 2008, ApJ, 689, 17 

\noindent Hopkins, P. F., Bundy, K., Murray N., Quataert E., Lauer T. R.,
Ma C.-P. 2009, MNRAS, 398, 898

\noindent Hopkins, P., et al. 2010b, ApJ, 715, 202

\noindent Khochfar S., Silk J. 2006a, ApJ, 648, L21 

\noindent Khochfar S., Silk J. 2006b, MNRAS, 370, 702 

\noindent Kurk J., et al., 2009, A\&A, 504, 331

\noindent La Barbera F., de Carvalho R. R., ApJ, 699, L76

\noindent Longhetti M., Saracco P., Severgnini P., et al., 2005, MNRAS, 361,
897

\noindent Longhetti M., Saracco P., Severgnini P., et al., 2007, MNRAS, 374, 614

\noindent Mancini, C., Daddi E., Renzini A., et al. 2010, MNRAS, 401, 933

\noindent McCarthy P. J., et al. 2004, ApJ, 614, L9

\noindent McGrath E., Stockton A., Canalizo G., Iye M., Maihara T. 2008,
ApJ, 682, 303

\noindent Muzzin A., van Dokkum P. G., Franx M., Marchesini D., Kriek M.,
Labb\'e I. 2009, ApJ, 706, L188

\noindent Naab T., Johansson P. H., Ostriker J. P., Efstathiou G. 2007, 
ApJ, 658, 710

\noindent Naab T., Johansson P. H., Ostriker J. P. 2009, ApJ, 699, L178

\noindent Newman A. B., Ellis R. S., Treu T., Bundy K. 2010, ApJ, 
[arXiv:1004.1331]

\noindent Onodera M. et al. 2010, ApJ, in press [arXiv:1004.2120]

\noindent Rettura A., et al., 2010, ApJ, 709, 512

\noindent Ryan Jr. E. R., et al. 2010, ApJ submitted [arXiv:1007.1460]

\noindent Saracco P., Longhetti M., Severgnini P., et al. 2003, A\&A, 398, 127

\noindent Saracco P., Longhetti M., Severgnini P., et al. 2005, MNRAS, 357, L40

\noindent Saracco P., Longhetti M., Andreon S., 2009, MNRAS, 392, 718

\noindent Saracco P., Longhetti M., Gargiulo A. 2010, MNRAS, 408, L21

\noindent Serra P., Trager S. C. 2007, MNRAS, 374, 769

\noindent Shen S., et al., 2003, MNRAS, 343, 978

\noindent Sommer-Larsen J., Toft S., 2010, ApJ, 721, 1755

\noindent Springel V., Hernquist L. 2005, 622, L9

\noindent Stark D. P., Bunker A. J., Ellis R. S., Eyles L. P., Lacy M. 2007,
ApJ, 659, 84 

\noindent Strazzullo V., et al. 2010, A\&A in press [arXiv:1009.1423]

\noindent Thomas D., Maraston C., Bender R., Mendes de Oliveira C. 2005,
ApJ, 621, 673

\noindent Thomas D., Maraston C., Schawinski K., Sarzi M., Silk J. 2010, 
MNRAS, 404, 1775

\noindent Tiret O., Salucci P., Bernardi M., Maraston C., Pforr J. 2010,
MNRAS, in press [arXiv:1009.5185]

\noindent Trujillo I., Feulner G., Goranova Y., et al. 2006, MNRAS, 373, L36

\noindent Trujillo I., Cenarro A. J., de Lorenzo-C\'aceres A., Vazdekis A.,
de la Rosa I. G., Cava A. 2009, ApJ, 692, L118

\noindent Valentinuzzi P., et al. 2010a, ApJ, 712, 226

\noindent Valentinuzzi P., et al. 2010b, ApJL, in press [arXiv:1007.4447]

\noindent van der Wel A., Bell E. F., van den Bosch F. C., Gallazzi A., Rix
H. 2009, ApJ, 698, 1232

\noindent van Dokkum P. G., et al. 2008, ApJ, 677, L5

\noindent van Dokkum P. G., Kriek M., Franx M. 2009, Nature, 460, 717

\noindent Wilkins S. M., Bunker A. J., Ellis R. S., Stark D., Stanway E. R.,
Chiu K., Lorenzoni S., Jarvis M. J. 2010, MNRAS, 403, 938

\noindent Wuyts S., et al. 2009, ApJ 700, 799
\label{lastpage}

\end{document}